\documentclass[showpacs,prc,epsfig]{revtex4}
\usepackage{graphicx}
\begin{document}
\title{Fermion-Boson Interactions and Quantum Algebras}
\vskip2cm
\author{A. Ballesteros}
\email{angelb@ubu.es}
\affiliation{Departamento de F\'{\i}sica,
Universidad de Burgos, Pza.~Misael Ba\~nuelos, E-09001 Burgos,
Spain}
\author{O. Civitarese}
\email{civitare@venus.fisica.unlp.edu.ar}
\affiliation{Departamento de F\'{\i}sica, Universidad Nacional de
La Plata, c.c.~67 1900, La Plata, Argentine}
\author{F.J.Herranz}
\email{fjherranz@ubu.es} \affiliation{Departamento de F\'{\i}sica,
Universidad de Burgos, Pza.~Misael Ba\~nuelos, E-09001 Burgos,
Spain}
\author{M. Reboiro}
\email{reboiro@venus.fisica.unlp.edu.ar} \affiliation{Departamento
de F\'{\i}sica, Universidad Nacional de La Plata, c.c.~67 1900, La
Plata, Argentine}
\date{\today}
\begin{abstract}
 Quantum Algebras (q-algebras) are used to describe interactions
between fermions and bosons. Particularly, the concept of a
$su_q(2)$ dynamical symmetry is invoked in order to reproduce the
ground state properties of systems of fermions and bosons
interacting via schematic forces. The structure of the proposed
$su_q(2)$ Hamiltonians, and the meaning of the corresponding
deformation parameters, are discussed.
\end{abstract}
\maketitle
\section{Introduction}

Group-theoretical methods have contributed significantly to the
study of the nuclear quantum-many-body problem. Classical examples
are the Lipkin model \cite{r1}, the Elliot SU(3) model \cite{r2},
the various realizations of Arima and Iachello Interacting Boson
Model (IBM) \cite{r3}, the Sch\"utte and Da Providencia  model
\cite{r4}, and the bi-fermion algebraic model of Geyer et al.
\cite{r5}, among other important contributions. We refer the
reader to the review article of Klein and Marshalek \cite{r6}, for
a comprehensive presentation of the problem.

Models based on the coupling between bi-fermions and bosons have
been introduced long ago \cite{r7,r8}. These models are
particularly suitable to describe condensation phenomena and
transitions from fermionic to bosonic phases. In particular, the model proposed
by Da Providencia and Sch\"utte (DPS) is a solvable model which
exhibits a phase transition between nucleonic and pionic
condensates \cite{r4}. Similar ideas have been applied to mock-up
the fundamentals of the non-perturbative, low energy, regime of
QCD \cite{r9,r10}. Recently, an extension of the Lipkin (LE) model
was proposed to take into account the interaction of pairs of
bi-fermions (quark-antiquark pairs) with external bosons (gluon
pairs) \cite{r11}. These constructions may help, as toy models, to
understand interactions between fermions and bosons in hadron
physics \cite{r13}. All these models share the non-perturbative
nature of the approximations \cite{r14}.

In addition to the conventional group theoretical approach, the
literature is rich in different realizations of deformed algebras
(q-algebras) whose mathematical foundations can be found, i.e., in
\cite{CP,r15,r16,GS}. Applications of concepts related to q-algebras to
some selected quantum mechanical examples can be found in
\cite{r17}. Recently published works on q-algebras, in connection
with quantum-many-body Hamiltonians \cite{r18,r19}, have shown the
suitability of the concept.

Several $q$-deformed versions of different schematic nuclear
models have been previously introduced through the substitution of
the symmetries of the original models by their $q$-analogues (see
\cite{r17} and references therein). An example of this ``direct
$q$-deformation" approach for a quantum optical fermion-boson
Hamiltonian was given in \cite{r20}. The phonon spectrum in $^4\mbox{He}$
has been also described by using a $q$-analogue of a $N$-oscillator Hamiltonian in
\cite{He4}.

Although the mathematical meaning of the deformation parameter (q)
which characterizes the commutation (anti-commutation) relations
between the generators of a given algebraic structure is
self-evident \cite{CP,r15,r16,GS}, the physical meaning of it is
less known. We recall that the exact $q$-deformed $su(2)$ symmetry
underlies the complete integrability of quantum systems like the
XXZ Heisenberg chain \cite{GS,r25,r26,r27} and Bloch electrons in
a magnetic field \cite{r28}. In nuclear models, except for the
quantum harmonic oscillator and the quantum rotor \cite{r17} where
$q$ may be interpreted as a stretching parameter of the
corresponding spectra, little is known about the potentiality of
q-algebras in more realistic cases.

With the above motivation in mind, and as a complementary tool to
standard non-perturbative many-body techniques \cite{r14}, in this
paper we propose the use of quantum algebras in the
construction of effective Hamiltonians.  We shall show that the
quantum algebra $su_q(2)$ \cite{r23} can be used to define, in a
natural way, new effective Hamiltonians which reproduce the same
ground state properties  and the spectrum of the ones based on
fermion-boson interactions. Advances in the same direction have been achieved in
quantum optics, where the interaction term of the
Dicke model has been described through
a $su_q(2)$ effective Hamiltonian \cite{r24}.

We would like to stress that such a quantum algebra approach
introduces a remarkable simplification of the models without any
significant loss of physical content. Explicitly, we shall show
how the DPS and LE Hamiltonians, originally defined on a
$su(2)\oplus h_3$ Lie algebra, where $su(2)$ is the algebra of
quasi-spin fermion operators and the Heisenberg algebra $h_3$
accounts for the boson degrees of freedom, can  be defined on the
$su_q(2)$ algebra alone. It is found that the new effective
$su_q(2)$-Hamiltonians reproduce accurately the physical
properties of the $su (2) \oplus h_3$ models, provided the
deformation parameter $q$ is suitably fitted in terms of physical
constraints.

We are confident that the present treatment can be successfully
applied to describe other physical systems, where the effective
motion is determined by the interaction between elementary fermion
and boson degrees of freedom. The correspondence between the
spectra of interacting fermion-boson systems and effective
q-deformed purely fermionic systems, which we demonstrate in this
work for some selected examples, may be a general feature common
to other fermion $\otimes$ boson systems. Such is the case, for
example, of nucleons interacting strongly {\it via} nuclear
$\lambda$-pole fields, which may also be represented as free
nucleons moving in a deformed central potential \cite{BMV2}.
Similarly, the non-perturbative domain of QCD, which is the theory
of the interactions between quark and gluons, may be viewed as an
effective theory of confined fermions \cite{r11,r13}, without
gluons. To summarize, this paper is devoted to the study of the
equivalence between systems of interacting fermions and bosons and
systems of q-deformed fermions. Clearly, we shall not deal with
the transformation of a given Hamiltonian onto a q-deformed space,
since this procedure leads, in general, to a completely different
Hamiltonian.

The paper is organized as follows. In section II we present the
basic aspects of the fermion-boson interaction models considered
in the work, and construct the associated deformed effective
Hamiltonians. In section III we discuss the behavior of the exact
solutions obtained for the different Hamiltonians. Conclusions are
drawn in section IV.

\section{Formalism}

In this section we shall briefly review the essentials of the DPS
and LE schematic models and discuss their realizations in the
framework of deformed algebras (hereon referred to as q-algebras).

\subsection{The DPS model.}

The DPS model \cite{r4} consists of $N=2\,\Omega$  fermions moving
in two single-shells. Each shell has a degeneracy $2\,\Omega$, and
its substates are labelled by the index $l=1,\dots,2\,\Omega$. The
energy-difference between shells is fixed by the energy scale
$\omega_f$. The creation and annihilation operators of particles
belonging to the upper level, are denoted by $a_{l}^{\dagger}$ and
$a_{l}$, respectively, while in the lower level, the creation and
annihilation operators for holes are denoted by $b_{l}^{\dagger}$
and $b_{l}$. The fermions are coupled to an external boson field
represented by the creation (annihilation) operators $B^{\dagger}$
($B$) and by the energy $\omega_b$, respectively. The DPS model
Hamiltonian can be interpreted as the one describing a system of
$N$ fermions (either nucleons or quarks), belonging to an
isospin-(flavor)-multiplet and $N$ spin projections (colors) in
interaction with bosons (either  pions or gluons), in a hadron
(QCD) scenario. The DPS Hamiltonian reads \cite{r4}

\begin{eqnarray}
H_{  } = \omega_f (T_0 +\Omega) +
          \omega _b B^{\dagger}B
+ G (T_{+} B^{\dagger}+T_{-} B), \label{aa}
\end{eqnarray}
where $G$ is the strength of the interaction in the particle-hole
channel.

The particle ($\nu$) and hole ($\overline{\nu}$) number operators are given by
\begin{equation}
\nu = \sum_{l} a_{l}^{\dagger}a_{l} ,  \;\;\;\;\; \overline{\nu }
= \sum_{l} b_{l}^{\dagger} b_{l}  , \label{ab}
\end{equation}
and the following bi-linear combinations of fermion operators
\begin{eqnarray}
T_{+} & = & \sum_{l} a_{l}^{\dagger}\,b_{l}^{\dagger}, \;\;\;
T_{-}  = (T_{+})^{\dagger}, \nonumber \\
T_{0} & = & \frac{1}{2}(\nu +\overline{\nu }) -\Omega ,
\label{ad}
\end{eqnarray}
are the generators of the $su(2)$ algebra:
\begin{equation}
[T_0,T_+]=T_+,\quad [T_0,T_-]=-T_-,\quad [T_+,T_-]=2 T_0 .
\label{ac}
\end{equation}

The Hamiltonian of Eq. (\ref{aa}) commutes with the operator
\begin{equation}
P = B^{\dagger} B- \frac {1}{2} (\nu+\overline {\nu}) =B^{\dagger} B-(T_0+\Omega).
\label{ae}
\end{equation}
Therefore, the matrix elements of $H$ can be calculated in a basis
labelled by the eigenvalues of the number operators for bosons and
fermions, as shown in \cite{r4},
\begin{equation}
| m_{\Omega},n \rangle = \sqrt {  \frac { (\Omega -m_{\Omega})! }{ (m_{\Omega}+\Omega)! ( 2 \Omega)!n!}  }
 T_+^{\Omega+m_{\Omega}}\,
 (B^{\dagger})^n |0\rangle .
\label{af}
\end{equation}
In this basis the eigenvalues of $P$ are given by
\begin{equation}
P | m_{\Omega},n \rangle = (n-m_{\Omega}-\Omega) | m_{\Omega},n \rangle.
\label{ag}
\end{equation}

In particular, we shall diagonalize $H$ in the subspace spanned by
the states $| m_{\Omega},L+m_{\Omega}+\Omega \rangle \equiv|
m_{\Omega};L,\Omega \rangle $ which have a fixed eigenvalue $L$ of
$P$
\begin{equation}
P | m_{\Omega};L,\Omega \rangle = L | m_{\Omega};L,\Omega \rangle.
\label{ag2}
\end{equation}
In this subspace, the non-zero matrix elements of $H$ are
\begin{eqnarray*}
\begin{array}{lll}
& \langle m_{\Omega};L,\Omega| H | m_{\Omega};L,\Omega \rangle =
\nonumber \\
& ~~~~~~~~~~~~~~~~~~~~~~~~~~~~
\omega_b L + (\omega_f+\omega_b) (\Omega+m_{\Omega}),  &
{\rm(9.a)}
\nonumber \\
& \langle m_{\Omega}+1;L,\Omega| H | m_{\Omega};L,\Omega \rangle  =  &
\nonumber \\
& ~~~ G \sqrt { (\Omega +m_{\Omega}+1) (\Omega-m_{\Omega})(L+\Omega+m_{\Omega}+1)}. & {\rm (9.b)}
\end{array}
\label{ah}
\end{eqnarray*}
\setcounter{equation}{9}
The dimension of the finite-dimensional subspace associated to each fixed eigenvalue
$L$, varies depending on the positive or the negative character of $L$.
For $L \geq 0$
the quantum number  $m_{\Omega}$ can take the values
\begin{eqnarray}
 m_{\Omega}=-\Omega, -\Omega+1,..., \Omega,
\label{ai}
\end{eqnarray}
and the Hilbert's subspace has dimension $2\Omega+1$.
In the case $L<0$, the values that $m_{\Omega}$ can take are
\begin{eqnarray}
m_{\Omega}=-L-\Omega,- L-\Omega+1,..., \Omega, \label{aj}
\end{eqnarray}
and accordingly, the dimension of the Hilbert's subspace is $2\,\Omega+L+1$.

\subsubsection{The DPS and effective $su_q(2)$ Hamiltonians}

The quantum algebra $su_q(2)$ is a Hopf algebra
deformation of $su(2)$ \cite{r23} whose generators are
${\tilde T}_\pm$ and ${\tilde T}_0$, and obey the commutation rules
\begin{equation}
[{\tilde T}_0,{\tilde T}_\pm]=\pm {\tilde T}_\pm,\quad  [{\tilde T}_+,{\tilde T}_-]=[2
{\tilde T}_0]_q.
\label{ak}
\end{equation}
Here the $q$-analogue $[x]_q$ of a given object $x$ (either a c-number
or an operator) is defined by
\begin{equation}
[x]_q  =   \frac{q^x-q^{-x}}{q-q^{-1}}    =   \frac {\sinh (z \; x)}{\sinh (z)}.
\label{al}
\end{equation}

Throughout the paper we shall use alternatively $q$ and $z$ (where
$q={\rm e}^z$) as the deformation parameter, furthermore we shall
assume that $q$ is real. Recall that the $su(2)$ algebra of Eq.
(\ref{ac}) is recovered from Eq. (\ref{ak}) in the limit $q\to 1$
($z\to 0$).

When $q$ is not a root of unity, the irreducible representations
of $su_q(2)$ are obtained as a straightforward generalization of
those of $su(2)$ \cite{CP,GS}. Namely,

\begin{eqnarray}
&& {\tilde{T}}_0 | j , m \rangle =m\,| j , m \rangle , \nonumber  \\
&& {\tilde{T}}_+ | j , m \rangle =\sqrt{ [j+m+1]_q [j-m ]_q}\,| j , m+1 \rangle ,
\nonumber \\ && {\tilde{T}}_- | j , m \rangle =\sqrt{ [j-m+1]_q [j+m ]_q}\,| j , m-1
\rangle .
\label{am}
\end{eqnarray}

The matrix elements of Eq. (9) correspond to a tridiagonal
finite dimensional matrix. Let us consider an effective
Hamiltonian, sharing the same property, which is defined as the
following function of the $su_q(2)$ generators
\begin{equation}
H_q = \omega_b L + (\omega_b + \omega_f ) ({\tilde{T}}_0 + \Omega)
+ \chi(q)  q^{  \frac {\tilde{T}_0}{2}}  ({\tilde{T}}_+
+{\tilde{T}}_-) q^{\frac {\tilde{T}_0}{2}} \label{an}
\end{equation}
where $\chi(q)$ is a scalar function and $H_q$ will be realized in
a $su_q(2)$ irreducible representation with the same dimension as
the subspace spanned by $| m_{\Omega};L,\Omega \rangle $
(therefore, with $j=j(\Omega, L)$). From Eq. (\ref{am}), the
non-vanishing matrix elements of Eq. (\ref{an}) read
\begin{eqnarray*}
&&\langle j , m |  H_q | j , m \rangle  = \omega_b L  +
(\omega_f+\omega_b) (m+\Omega) ,
 \label{16a} ~~~~~(16.a) \nonumber \\
& & \langle j , m+1 |  H_q | j , m \rangle  =
 \nonumber \\ & & \;\;\;\;\;\;\;\;\;\;\;\;\;\;
\chi(q) {q}^{(m+\frac12)}\sqrt{ [j+m+1]_q [j-m ]_q}. \label{16b}
~~~~(16.b)
\end{eqnarray*}
\setcounter{equation}{16}
In order to fit the dimensions, in the previous equations we take
$j=\Omega$ and $m=m_{\Omega}$  for the effective $L \ge 0$ model,
while for $L<0$, $j=\Omega+ \frac L2$ and $m=m_{\Omega}+\frac L2$.

Note that, apparently, the Hamiltonians $H$ of Eq. (\ref{aa}) and
$H_q$ of Eq. (\ref{an}) seem to be  quite different, since the
latter has no bosonic degrees of freedom.  In fact, the
non-deformed limit $q\to 1$ of Eq. (\ref{an}), is the non-deformed
$su(2)$ Hamiltonian
\begin{equation}
H = \omega_b L + (\omega_b + \omega_f ) ({{T}}_0 + \Omega) +
\chi(1) (T_++T_-) . \label{ao}
\end{equation}
which cannot be obtained from Eq. (\ref{aa}) through any
transformation. The main result of this procedure is that the
bosonic degrees of freedom included in Eq. (\ref{aa}) may be
absorbed by the $q$-deformation in Eq. (\ref{an}) provided that
$q$ is defined as an appropriate function of both $\Omega$ and
$L$, a trade-off leading to the purely fermionic structure of Eq.
(\ref{an}). In this way it is possible to regard $H_q$ as an
effective Hamiltonian with physical properties similar to those of
$H$. In particular, we shall determine numerically the optimal
values of the deformation parameter $q$ by imposing that the
spectrum of the $q$-Hamiltonian of Eq.~(\ref{an}) be as close as
possible to that of Eq.~(\ref{aa}). In so doing, the function
$\chi(q)$ has been chosen as
\begin{equation}
\chi(q)= \frac {G \sqrt{ L+\Omega+m_{\Omega}^0+1 }
\sqrt{(\Omega+m_{\Omega}^0+1) (\Omega-m_{\Omega}^0)}} {
{q}^{(m^0+\frac12)}  \sqrt{ [j+m^0+1]_q [j-m^0 ]_q}  }, \label{aq}
\end{equation}
where $m_{\Omega}^0$ is the value of $m_{\Omega}$ that maximizes
the matrix element of Eq. (9.b), and $m^0=m_\Omega^0$ for $L \geq
0$, while $m^0=m_\Omega^0+ \frac L2$ for $L<0$. This choice
ensures that the maximum values of the interaction terms of the
Hamiltonians $H$ and $H_q$  coincide (see \cite{r24}), as it is
shown in Section III.

The main role of the exponentials $q^{\frac {\tilde{T}_0}{2}}$ in
Eq. (\ref{an}) is to break the $m\leftrightarrow -m$ symmetry of
the effective model, since this is one of the main effects of the
nonlinearity introduced by the fermion-boson coupling in  Eq.
(\ref{aa}). This effect could be reproduced, also, through
functions others than exponentials of the $\tilde{T}_0$ operator.
The effective fermionic Hamiltonian could also be defined by using
more involved functions on the non-deformed $su(2)$ algebra, since
the main constrain is the block-structure of  Eq. (\ref{aa}).
Nevertheless, we would like to stress that the essential advantage
of using both the $su_q(2)$ operators of Eq. (\ref{am}) and the
exponential form of the effective Hamiltonian of Eq. (\ref{an}) is
that the eigenvalues of the interaction term
\begin{equation}
H_q^{\mbox{int}}=q^{ \frac {\tilde{T}_0}{2}} ({\tilde{T}}_+
+{\tilde{T}}_-)
  q^{\frac {\tilde{T}_0}{2}}
\label{hqint}
\end{equation}
are just the $q$-numbers $[2\,m]_q$ ($m=-j,\dots,+j$), and its
eigenvectors are known in analytic form. They are related to
$q$-Krawtchouk polynomials \cite{r24}. Therefore, the results here
presented seem to indicate that certain interactions between
fermions and bosons can be accurately described by using
$q$-fermions as quasiparticles (i.e; effective fermionic degrees
of freedom) under the exactly solvable interaction given by the
Hamiltonian $H_q^{\mbox{int}}$.

\subsection{Extended Lipkin models.}

As a second example of Hamiltonians including fermionic and
bosonic degrees of freedom, let us introduce the Lipkin-type
Hamiltonian
\begin{eqnarray}
H_{  } =  \omega_f (T_0 +\Omega) +
          \omega_b B^{\dagger}B
+ G (T_{+}^2 B+T_{-}^2 B^{\dagger}).
\label{b1a}
\end{eqnarray}
The fermion sector of the model is described by two levels, with
energies $\pm \frac {\omega_f}{2}$ and degeneracies  $2 \Omega$.
The fermions interact with bosons of energy $\omega_b$.

The Hamiltonian of Eq.~(\ref{b1a}) commutes with the operator
\cite{twobosons}

\begin{equation}
P_{(+)}  = B^{\dagger} B+ \frac {1}{2} (T_0 +\Omega) ,
\label{b1b}
\end{equation}
therefore, the matrix elements of $H$ can be calculated in a basis
labelled by the eigenvalues of $P$.

If we consider   the basis of Eq. (\ref{af}), we find that the
eigenvalues of $P_{(+)} $  are given by
\begin{equation}
P_{(+)}  | m_{\Omega},n \rangle =
(n+ \frac 12 (\Omega+m_{\Omega})) | m_{\Omega}, n \rangle,
\label{b1c}
\end{equation}
we shall consider the subspace spanned by the states with $L$
fixed $| m_{\Omega},L-\frac 12 (\Omega+m_{\Omega}) \rangle \equiv|
m_{\Omega};L,\Omega \rangle $.

Hence the non-zero matrix elements of $H$ read
\begin{eqnarray*}
\langle m_{\Omega};L,\Omega| H | m_{\Omega};L,\Omega \rangle =
\omega_b L + (\omega_f -\frac 12 \omega_b)(\Omega+m_{\Omega}), \\
(23.a)  \\
\langle m_{\Omega}+2;L,\Omega| H | m_{\Omega};L,\Omega \rangle =
G \sqrt {L- \frac 12 (\Omega+m_{\Omega})} \times  \\
\sqrt {(\Omega+m_{\Omega}+2) (\Omega+m_{\Omega}+1) (\Omega-m_{\Omega})
(\Omega-m_{\Omega}-1) }, \\
(23.b)
\label{b1d}
\end{eqnarray*}
\setcounter{equation}{23}
with
\begin{eqnarray}
&& L \ge \Omega, L \; {\rm integer}, \qquad
m_{\Omega}+\Omega=0,2,..., 2\Omega,
\nonumber \\
&& L > \Omega, L \; {\rm half-integer}, \nonumber \\
&& \,\,\,\,\,\,\,\,\,\,\,\,\,\,\,\,\,\,\,\,\, \,
m_{\Omega}+\Omega=1,3,..., 2\Omega-1,
\nonumber \\
&& L < \Omega, L \; {\rm integer} , \quad m_{\Omega}+\Omega=
0,2,...,2 L,
\nonumber \\
&& L < \Omega, L \; {\rm half-integer} , \nonumber \\
&& \,\,\,\,\,\,\,\,\,\,\,\,\,\,\,\,\,\,\,\,\, m_{\Omega}+\Omega=
0,2,...,2L-1. \nonumber \\
\label{b1e}
\end{eqnarray}

The Hamiltonian

\begin{eqnarray}
H_{  } = \omega_f (T_0 +\Omega) +
          \omega_b B^{\dagger}B
+ G (T_{+}^2 B^{\dagger}+T_{-}^2 B),
\label{b2a}
\end{eqnarray}
is another Lipkin-type Hamiltonian, which differs from Eq.
(\ref{b1a}) in the ground state correlations \cite{r11}. Since it
commutes with the operator

\begin{equation}
P_{(-)} = B^{\dagger} B- \frac {1}{2} (T_0 +\Omega),
\label{b2b}
\end{equation}
its matrix elements can be calculated in a basis labelled by the
eigenvalues of $P_{(-)}$. Once again this basis is just Eq.
(\ref{af}), where the eigenvalues of $P_{(-)}$ read
\begin{equation}
P_{(-)}  | m_{\Omega},n \rangle = (n- \frac 12
(\Omega+m_{\Omega})) | m_{\Omega},n \rangle, \label{b1c1}
\end{equation}
and we shall compute the matrix elements in the subspace spanned by
the states
$| m_{\Omega},L+\frac 12 (\Omega+m_{\Omega}) \rangle \equiv| m_{\Omega};L,\Omega \rangle $.
The non-zero matrix elements of $H$  are given by
\begin{eqnarray*}
\langle m_{\Omega};L,\Omega| H | m_{\Omega};L,\Omega \rangle
= \omega_b L + (\omega_f+ \frac 12 \omega_b)(\Omega +m_{\Omega}) , \\
(28.a) \\
\langle m_{\Omega}+2;L,\Omega| H | m_{\Omega};L,\Omega \rangle  =
G \sqrt {L+ \frac 12 (\Omega+m_{\Omega})+1} \times \\
\;\;\;\;\;\;
\sqrt { (\Omega+m_{\Omega}+2) (\Omega+m_{\Omega}+1) (\Omega-m_{\Omega}) (\Omega-m_{\Omega}-1) }, \\
(28.b)
\label{b2d}
\end{eqnarray*}
\setcounter{equation}{28}
where the dimension of the subspace depends on $L$ and $\Omega$ in the form
\begin{eqnarray}
&& L \ge 0, L \; {\rm integer}, \quad m_{\Omega}+\Omega=0,2,..., 2\Omega, \nonumber \\
&& L > 0, L \; {\rm half-integer}, \nonumber \\
&& \,\,\,\,\,\,\,\,\,\,\,\,\,\,\,\,\,\,\,\,\,
 m_{\Omega}+\Omega=1,3,..., 2\Omega-1, \nonumber \\
&& L < 0, L \; {\rm integer} ,\quad   m_{\Omega}+\Omega=-2 L ,-2
L+2,..., 2\Omega, \nonumber
\\ && L < 0, L \; {\rm half-integer} , \nonumber \\
&& \,\,\,\,\,\,\,\,\,\,\,\,\,\,\,\,\,\,\,\,\,
m_{\Omega}+\Omega=-2 L,-2 L+2,...,2\Omega-1. \nonumber \\
\label{b2e}
\end{eqnarray}

\subsubsection{The extended Lipkin and $su_q(2)$ effective Hamiltonians.}

As for the case of the DPS  model, we introduce  an effective
Hamiltonian for Eq. (\ref{b1a}), which is again defined on the
$su_q(2)$ generators,
\begin{equation}
H_q = \omega_b L + ( \omega_f - \frac 12 \omega_b ) ( {\tilde{T}}_0 + \Omega )
+ \chi(q) q^{{\tilde{T}}_0}  ({\tilde{T}}_+^2 +{\tilde{T}}_-^2) q^{{\tilde{T}}_0}.
\label{ca}
\end{equation}
This Hamiltonian has the following non-vanishing matrix elements
\begin{eqnarray*}
\langle j , m |  H_q | j , m \rangle   =
\omega_b L + (\omega_f -\frac 12 \omega_b) (\Omega+m), ~~~~~~ \\
(31.a) \\
\langle j , m+2 |  H_q | j , m \rangle = \chi(q) {\rm q}^{2 (m+1)} \times ~~~~~~~~~~~~~~~~~~ \\
\;\;\;  \sqrt { [j+m+2]_q [j+m+1]_q [j-m]_q [j-m-1]_q }. \\
(31.b) \\
\label{cb}
\end{eqnarray*}
\setcounter{equation}{31}
As a the first step in order to fit the dimensions of Eq.
(\ref{b1a}) and of Eq. (\ref{ca}) we have to find the appropriate
relation $j=j(\Omega,L)$ and, as a consequence,
$m=m(m_\Omega,\Omega,L)$. Afterwards, we consider as the effective
Hamiltonian the restriction of the matrix elements of Eq.
(31) to the invariant subspace spanned by $| j , m \rangle$
with $m=-j, -j+2,\dots, j-2, j$. In this way we obtain the
effective matrix elements
\begin{equation}
\langle j , m+2 |  H_q | j , m \rangle  = \chi(q) \,h(L, \Omega, m_\Omega).
\end{equation}

For values of $L\ge \Omega$ ($L$ integer), we find $j=\Omega$, $m=m_\Omega$ and the
function $h(L, \Omega, m_\Omega)$ is
\begin{eqnarray}
& & h(L, \Omega, m_\Omega)= {\rm q}^{2(m_\Omega+1)} \times
\nonumber \\
\,\,\, & &
\sqrt { [ {\Omega}+m_\Omega+2 ]_q [ {\Omega}+m_\Omega+1 ]_q [ {\Omega}- m_\Omega ]_q [{\Omega}-m_\Omega-1 ]_q  }.
\nonumber \\ \label{cc}
\end{eqnarray}
When $L < \Omega$ ($L$ integer), we have that $j=L+\frac 12$,
$m=m_\Omega+\Omega-L-\frac 12$ and
\begin{eqnarray}
& & h(L, \Omega, m_\Omega)= {\rm q}^{2 (m_\Omega + \Omega -L+ \frac 12 )} \times \nonumber \\
& & \,\,\,\,\, \sqrt { [ {\Omega}+m_\Omega-2 L ]_q [{\Omega}+m_\Omega-2 L-1 ]_q } \times \nonumber \\
& &  \,\,\,\,\,
\sqrt{[ {\Omega}+m_\Omega+1 ]_q [ {\Omega}+m_\Omega+2 ]_q  }.\nonumber \\
\label{cd}
\end{eqnarray}

The function $\chi(q)$ is defined by
\begin{eqnarray}
& &\chi(q)  = G \sqrt { L- \frac 12 (\Omega+m_{\Omega}^0)} \times
\nonumber \\
& & \;\;
\frac
{ \sqrt { (\Omega+m_{\Omega}^0+2) (\Omega+m_{\Omega}^0+1)  (\Omega-m_{\Omega}^0) (\Omega-m_{\Omega}^0-1)} }
{h(L,\Omega,m_{\Omega}^0)},
\nonumber \\
\label{ce}
\end{eqnarray}
where $m_{\Omega}^0$ is chosen as the value of $m_\Omega$ that maximizes Eq. (23.b).
Following the arguments presented in subsection A we shall search
for values of the $q$-dependent coupling (of Eq. (\ref{ca}), Eq.
(\ref{ce})) which may absorb bosonic degrees of freedom of Eq.
(\ref{b1a}) and yields a comparable spectrum for the purely
fermionic q-deformed Hamiltonian of Eq. (\ref{ca}).

Similarly, for the Hamiltonian of Eq. (\ref{b2a}), we can write
the effective $su_q(2)$ coupling

\begin{equation}
H_q = \omega_b L + (\omega_f + \frac 12 \omega_b )
({\tilde{T}}_0 + \Omega)
 +  \chi(q) q^{{\tilde{T}}_0}  ({\tilde{T}}_+^2 +{\tilde{T}}_-^2) q^{{\tilde{T}}_0}.
\label{cf}
\end{equation}

Now the matrix elements of $H_q$ are given by
\begin{eqnarray*}
\langle j , m |  H_q | j , m \rangle = \omega_b L +
(\omega_f+ \frac 12 \omega_b)(\Omega + m), ~~~~~~~~~~~~ \\
(37.a) \\
\langle j , m+2 |  H_q | j , m \rangle =\chi(q) {\rm q}^{2 (m+1)} \times ~~~~~~~~~~~~~~~~~~~~~~~ \\
\;\;\; \sqrt { [j+m+2]_q [j+m+1]_q [j-m]_q [j-m-1]_q }. \\
(37.b) \\
\label{cg}
\end{eqnarray*}
\setcounter{equation}{37}
Note that, the Hamiltonians of Eq. (\ref{ca}) and Eq. (\ref{cf})
differ in the unperturbed sector, and that once again we have to
fit the dimension of the $su_q(2)$ operator through the
appropriate choice of the quantum numbers $j$ and $m$. For values
of $L\ge 0$ ($L$ integer), we find $j=\Omega$ and $m=m_\Omega$. In
the case $L < 0$ ($L$ integer), we have $j=L+\Omega+ \frac 12$ and
$m=m_\Omega+L+\frac 12$. We recall that the effective $su_q(2)$
matrix is given by the matrix elements of Eq. (37) computed
within the subspace spanned by $| j , m \rangle$ where $m=-j,
-j+2,\dots, j-2, j$.

Finally, the adopted expression for $\chi(q)$, in Eq. (\ref{cf}) is
\begin{eqnarray}
& &\chi(q)  = G {\rm q}^{-2(m^0+1)} \;
\sqrt { L+ \frac 12 (\Omega+m_{\Omega}^0)+1} \times
\nonumber \\
& & \;\; \sqrt { \frac
{ (\Omega+m_{\Omega}^0+2) (\Omega+m_{\Omega}^0+1)
  (\Omega-m_{\Omega}^0) (\Omega-m_{\Omega}^0-1) }
{ [ j+m^0+2 ]_q [ j+m^0+1 ]_q [ j-m^0 ]_q [j-m^0-1 ]_q }
},
\nonumber \\
\label{ch}
\end{eqnarray}
and  $m_{\Omega}^0$ is chosen as the value of $m_\Omega$ that maximizes Eq. (28.b),
accordingly, $m^0=m_\Omega^0$ for $L \geq 0$, and $m^0=m_\Omega^0+
L+ \frac 12 $ for $L<0$.

Before ending with this section, we shall summarize the main steps
of the above formalism. We have written, for different fermion
$\otimes$ boson Hamiltonians, q-deformed purely-fermionic
Hamiltonians where the information about boson degrees of freedom
is absorbed in the definition of the q-dependent strength
$\chi(q)$. The actual value of $\chi(q)$ depends on the
deformation parameter, which may be determined from the comparison
between the spectra of the fermion $\otimes$ boson and q-deformed
fermion Hamiltonians. We shall discuss the feasibility of this
procedure in the next section III.

\section{Results and Discussion}

We have calculated the spectra of the Hamiltonians introduced in
the previous section. The calculations have been performed by
fixing the following set of parameters: $\omega_f=\omega_b=1$, in
arbitrary units of energy, and for $N=2 \Omega= 30$ particles,
unless stated. The parameter $x= G \sqrt {  \frac { 2
\Omega}{\omega_f \omega_b} }$, was taken as the dimensionless
coupling between fermions and bosons, for the case of Hamiltonian
of Eq. (\ref{aa}), and it is defined as $x= G \frac {4
\Omega}{\sqrt{\omega_f \omega_b}}$, for the case of the
Hamiltonians of Eqs. (\ref{b1a}) and (\ref{b2a}). As we shall
discuss later on, actual values of $x$ are indicative of the phase
preferred by the system (in the sense of the dominance of the
fermionic or bosonic degrees of freedom on the structure of the
ground state) \cite{twobosons}. In general, we shall talk of a
{\it normal} phase, of any of the fermion $\otimes$ boson
Hamiltonians of the previous section, when the correlated ground
state is the eigenstate of the symmetry operator $P$ with the
eigenvalue $L=0$. The denomination {\it deformed} phase will be
assigned to cases where the correlated ground state is an
eigenstate of $P$ with eigenvalue $L \neq 0$. The bosonic or
fermionic structure of the deformed phase is determined by the
sign of $L$, following the corresponding definition of $P$.

Let us start with the DPS model. The coupling $x=0.5$ yields a
normal solution of the DPS Hamiltonian. The value $x=1.5$ is
consistent with a deformed solution of it. Figure 1, cases (a) and
(c), shows the evolution of the ground state upon $L$. In the same
figure we present the results of the $q$-deformed Hamiltonian
corresponding to the DPS Hamiltonian. Figure 1, cases (b) and (d),
shows the behavior of the deformed parameter $z=\ln (q)$, as a
function of $L$, which reproduces the ground state energies of the
insets (a) and (c). The values of $z$ have been chosen so that the
ground states energies of the DPS model and the ones of the
$su_q(2)$ effective model of Eq. (\ref{an}) differ in less than $1
\%$. Figure 2 shows the evolution of the values of $L$, $z$ and
$\chi(q)$, at the absolute ground state, for different values of
the coupling constant $x$. Two different phases can be identified,
depending on the value of $x$. The normal phase corresponds to
values of $x \le 1$, with $L=0$ and $z$ almost constant, and the
deformed phase corresponds to values of $x >1$, with values of $L
> 0$ and decreasing values of $z$.

Figure 3 displays the comparison between the matrix elements of
the DPS model and the ones obtained with the effective hamiltonian
of Eq.~(\ref{an}). The scaling of Eq. (18) was performed as
indicated in the text. These results support nicely the adopted
procedure, since the agreement between both set of matrix elements
is rather acceptable.

Figure 4 shows the results of the integrated hamiltonian
energy-density, corresponding to the hamiltonias of Eq. (1) and
Eq. (15). Again in this case the agreement between both set of
results was verified within the computer accuracy.

The above results, shown in Figures (1)-(4), demonstrate that both
the ground state energy and the spectrum of the DPS model can be
represented by the effective $su_q(2)$ Hamiltonian of
Eq.~(\ref{an}), by fixing the value of $z(q)$, which is the
parameter related with the q-deformation.

A similar analysis can be performed for the LE models of
subsection II.B. Figure 5 represents the ground state energy of
the fermion $\otimes$ boson Hamiltonian of Eq.~(\ref{b1a}), and
the behavior of the parameter $z$ of the corresponding q-deformed
version, Eq.~(\ref{ca}). Also, in the same figure, the ground
state energy of the q-deformed Hamiltonian (Eq.~(\ref{ca})) is
given as a function of $L$. The insets (a) and (b), of Figure 5,
show the results corresponding to $x=0.5$ (normal phase), while
insets (c) and (d) show the results obtained with $x=1.5$
(deformed phase). As for the case of the DPS model, we have chosen
$z$ so that the ground state energy of the hamiltonian of
Eq.~(\ref{b1a}) and that of Eq.~(\ref{ca}) coincide within $1 \%$,
for each value of $L$.

Figure 6 displays the behavior of $L$, $z$ and $\chi(q)$, at the
absolute ground state energy, for different values of the coupling
constant $x$, for the Hamiltonians of Eqs. (\ref{b1a}) and
(\ref{ca}). As for the case of the DPS model, the results shown in
this figure correspond to two different phases, which can be
identified by the value of $x$. Similarly to the case of Figure 2,
the normal phase corresponds to $x \le 1$, $L=0$ and $z(q)$ nearly
constant. The deformed phase corresponds to $x > 1$, $L \neq 0$
and increasing values of $z(q)$.

Figure 7 displays the comparison between the spectrum of the
Hamiltonian of Eq.~(\ref{b1a}) and the spectrum of the effective
Hamiltonian of Eq.~(\ref{ca}). As done for the cases of Eq. (1)
and Eq.(15) (see Figure 4) we have calculated the integrated
hamiltonian energy-density  (number of eigenvalues per unit
energy-interval). Also in this case, the scaling procedure yields
almost identical results, within computer accuracy, as compared to
the original hamiltonian.

Finally, the ground state energies, the q-deformation parameter,
the q-depending coupling, and the comparison between the spectra,
for the case of the Hamiltonians of Eqs. (\ref{b2a}) and
(\ref{cf}) are shown in Figures 8-10, respectively. From the
results shown in Figure 8, cases (b) and (d), it is seen that
there is a particular value of $L$, for which $z(q) \approx 0$. It
means that, for this particular value of $z(q)$, the $su(2)$
symmetry is dynamically restored. Figure 9 shows the behavior of
$L$ (inset (a)), $z(q)$ (inset (b)) and $\chi(q)$ (inset (c)),
taken at the absolute ground state energy, as a function of $x$.
Figure 10 displays the comparison between the spectrum of the
Hamiltonian of Eq.~(\ref{b2a}) and the one obtained with the
effective hamiltonian of Eq.~(\ref{cf}), with $z=0.00440$.

A systematic feature emerges from the above discussed series of
results and it is related with the replacement of the boson
degrees of freedom, which are present in the considered initial
Hamiltonians, by the effective q-dependent coupling. In the three
cases, which we have considered, the spectrum of the fermion
$\otimes$ boson system and the spectrum of the q-deformed purely
fermionic system agree, for certain non-trivial values of the
q-deformation parameter $z(q)$. The procedure works reasonably,
for the rotor-like structure of the DPS Hamiltonian, as well as
for the vibrational-like structure of the LE Hamiltonians. There
is a trend in the dependence of $z(q)$ upon $L$, which is the
parameter associated to the symmetry in the fermion $\otimes$
boson space. It is symmetric for the case of the DPS Hamiltonian
and almost asymmetric for the case of the Lipkin Hamiltonians.
Also, $z(q)$ resembles more the behavior of an order parameter,
for the case of the Lipkin Hamiltonians, than for the DPS one.
Concerning relatives values of $z(q)$, the q-deformed versions of
the  Hamiltonians of Eqs.~(\ref{aa}) and (\ref{b1a}), required
values of $z(q)$ ($\approx 0.03$) which are larger than the value
of $z(q)$ corresponding to the q-deformed version of Hamiltonian
of Eq.~(\ref{b2a}). This result shows the sensitivity of the
chosen value of $z$ upon the vibrational or rotational-like
character of the fermion $\otimes$ boson picture.

\section{Conclusions}

In this work we have shown that effective $su_q(2)$ Hamiltonians
can be introduced in order to reproduce the ground state
properties  and the spectrum of different interacting
fermion-boson Hamiltonians. In this respect, the bosonic part of
the interactions can be effectively embedded as an appropriate
q-deformation of the $su(2)$ fermionic algebra.

The results presented at this work show the existence of a close
relation between the deformation parameter, $z(q)$, which fixes
the strength $\chi(q)$ of the purely fermionic q-deformed
Hamiltonians, and the eigenvalue, $L$, of the symmetry operator
$P$, associated to the fermion $\otimes$ boson Hamiltonians. Both
$z(q)$, in the case of the $su_q(2)$ effective models, and $L$,
for the fermion-boson interactions, display a critical behavior as
a function of the coupling constant $x$.

Because of the relevance of the DPS model in the description of
hadronic systems \cite{bes} and the nice agreement obtained with
the q-deformed version of it, we are confident about the
potentiality of q-deformed representations in more involved
physical scenarios.

Work is in progress concerning the extension of the presented
formalism  to non-perturbative QCD.

\section*{Acknowledgements}

This work has been partially supported by the Ministerio de
Ciencia y Tecnolog\'{\i}a (Espa\~na) under Project BFM2000-1055
and by the CONICET (Argentina). M.R. is grateful to the
Universidad de Burgos  for hospitality. (M.R.) acknowledge
financial support of the Fundacion Antorchas and of Universidad de
Burgos (Invited Professors Program).

\begin{figure}
\begin{center}
\includegraphics[width=10cm]{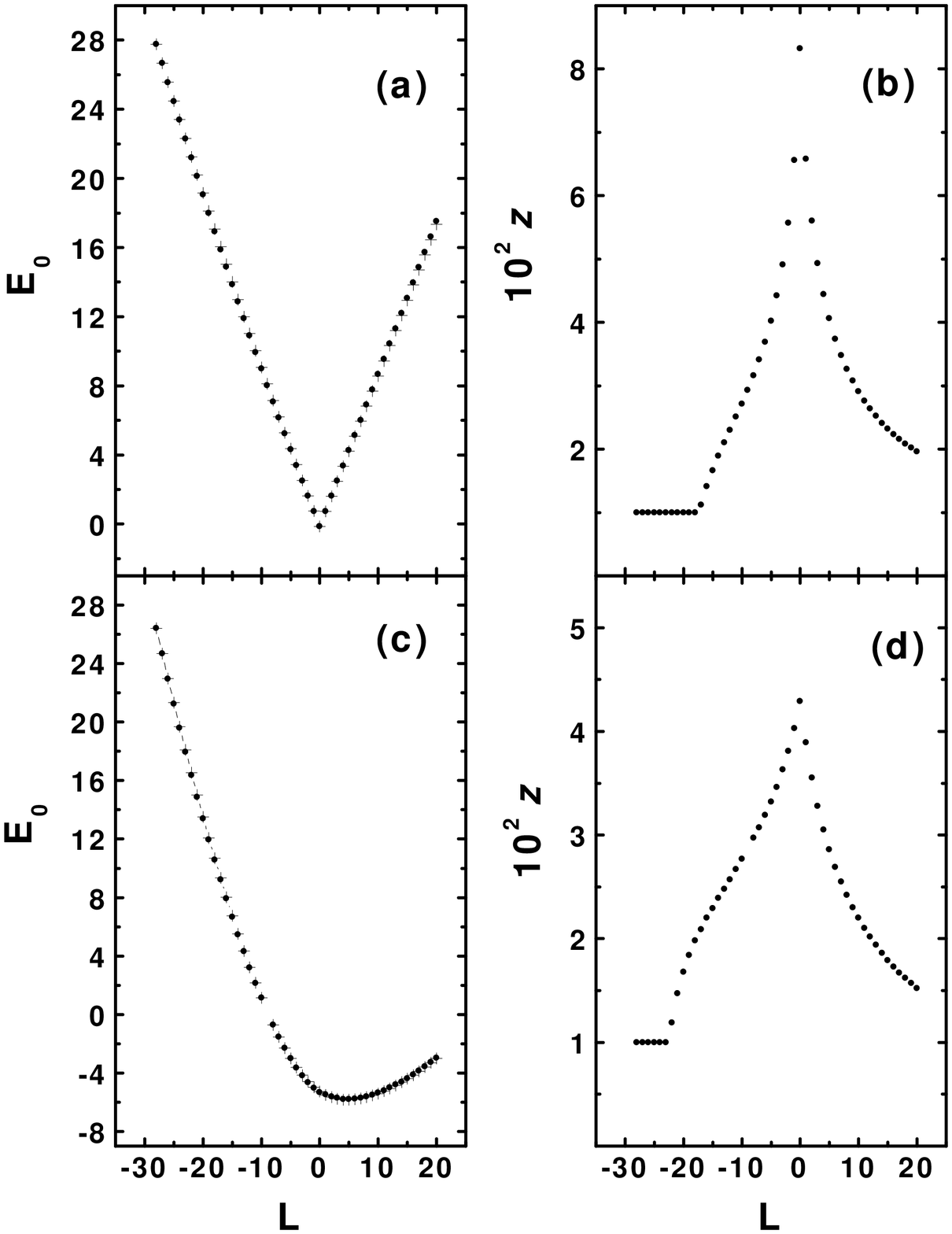}
\end{center}
\caption{Ground-state energy, $E_0$, in arbitrary units,  and
deformation parameter, $z=\ln (q)$, as a function of $L$. Insets
(a) and (b) show results for the case $N=2 \Omega=30$,
$\omega_f=1$, $\omega_b=1$, and $x=G \sqrt{\frac{2
\Omega}{\omega_f \omega_b}}=0.5$, while insets (c) and (d)
correspond to $x=1.5$. The exact ground state energy corresponding
to the DPS model of Eq. (\ref{aa}) is denoted by crosses while the
one corresponding to the effective $su_q(2)$ model of Eq.
(\ref{an}) is denoted by circles.} \label{fig:fig1}
\end{figure}

\begin{figure}
\begin{center}
\includegraphics[width=10cm]{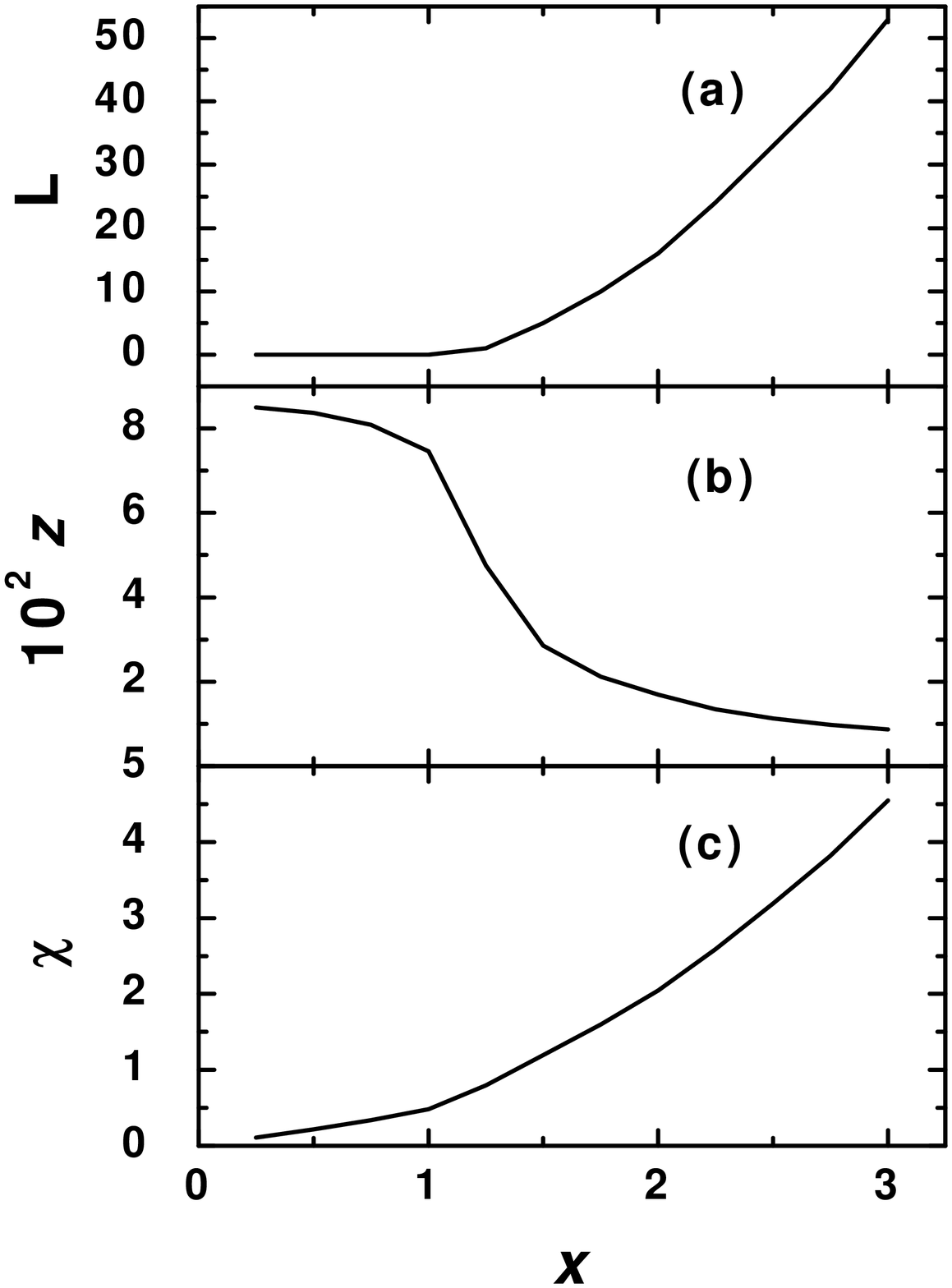}
\end{center}
\caption{Values of $L$ (inset (a)), $z$ (inset (b)) and $\chi$
(inset (c)), at the absolute ground state energy, as a function of
$x$. The figure displays the results corresponding to the case
$N=30$, $\omega_f=1$ and $\omega_b=1$ for different values of $x$,
and for the Hamiltonians~ of Eq. (\ref{aa}) and (\ref{an})(see
Figure 1).} \label{fig:fig2}
\end{figure}

\begin{figure}
\begin{center}
\includegraphics[width=10cm]{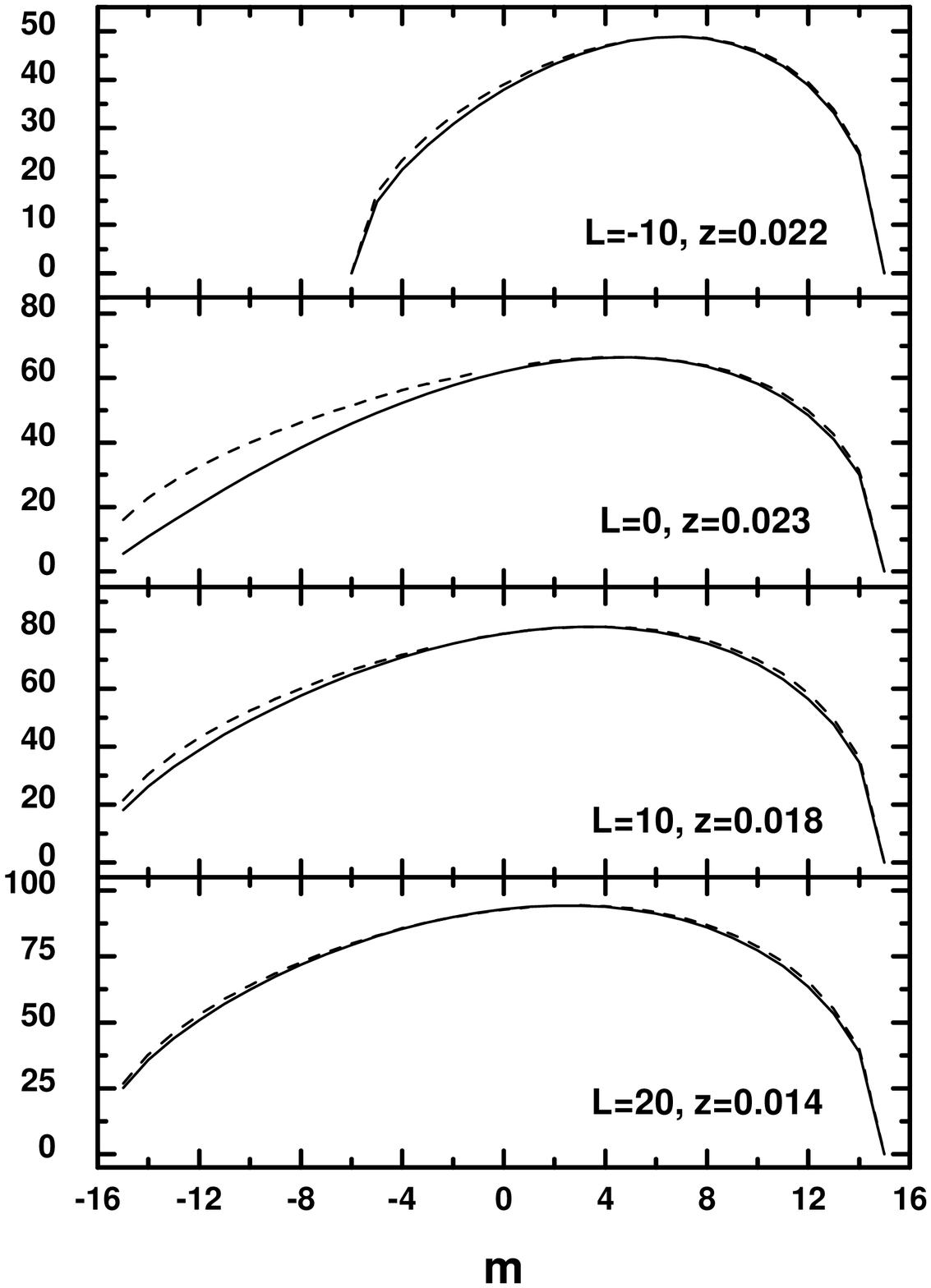}
\end{center}
\caption{Scaling procedure of Eq. (18). The matrix elements of Eq.
(9.b)(solid line) and Eq. (16.b)(dashed line) are shown as a
function of the $m$-quantum number. The values of $L$ and $z$ are
indicated in the insets. } \label{fig:fig3}
\end{figure}

\begin{figure}
\begin{center}
\includegraphics[width=10cm]{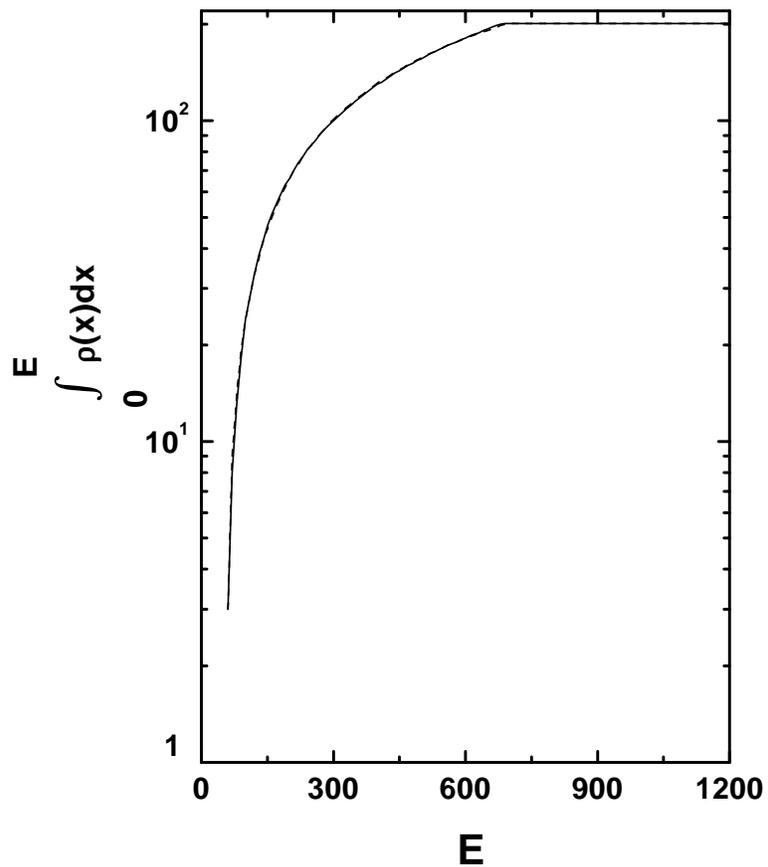}
\end{center}
\caption{Integrated density of states, as a function of the
energy, corresponding to the Hamiltonians of Eq.~(\ref{aa}) (solid
line) and of Eq.~(\ref{an}) (dashed line). Both results coincide
in the curve shown in the figure. The calculations were performed
for $N=200$, $\omega_f=1$, $\omega_b=1$, and $x=1.5$. For the DPS
model, Eq.~(\ref{aa}), the value $L=34$ was used. The spectrum of
the effective $su_q(2)$ hamiltonian of Eq.~(\ref{an}) was
calculated with $z=0.004$. Both curves coincide within the
resolution of the diagram} \label{fig:fig4}
\end{figure}
\begin{figure}
\begin{center}
\includegraphics[width=10cm]{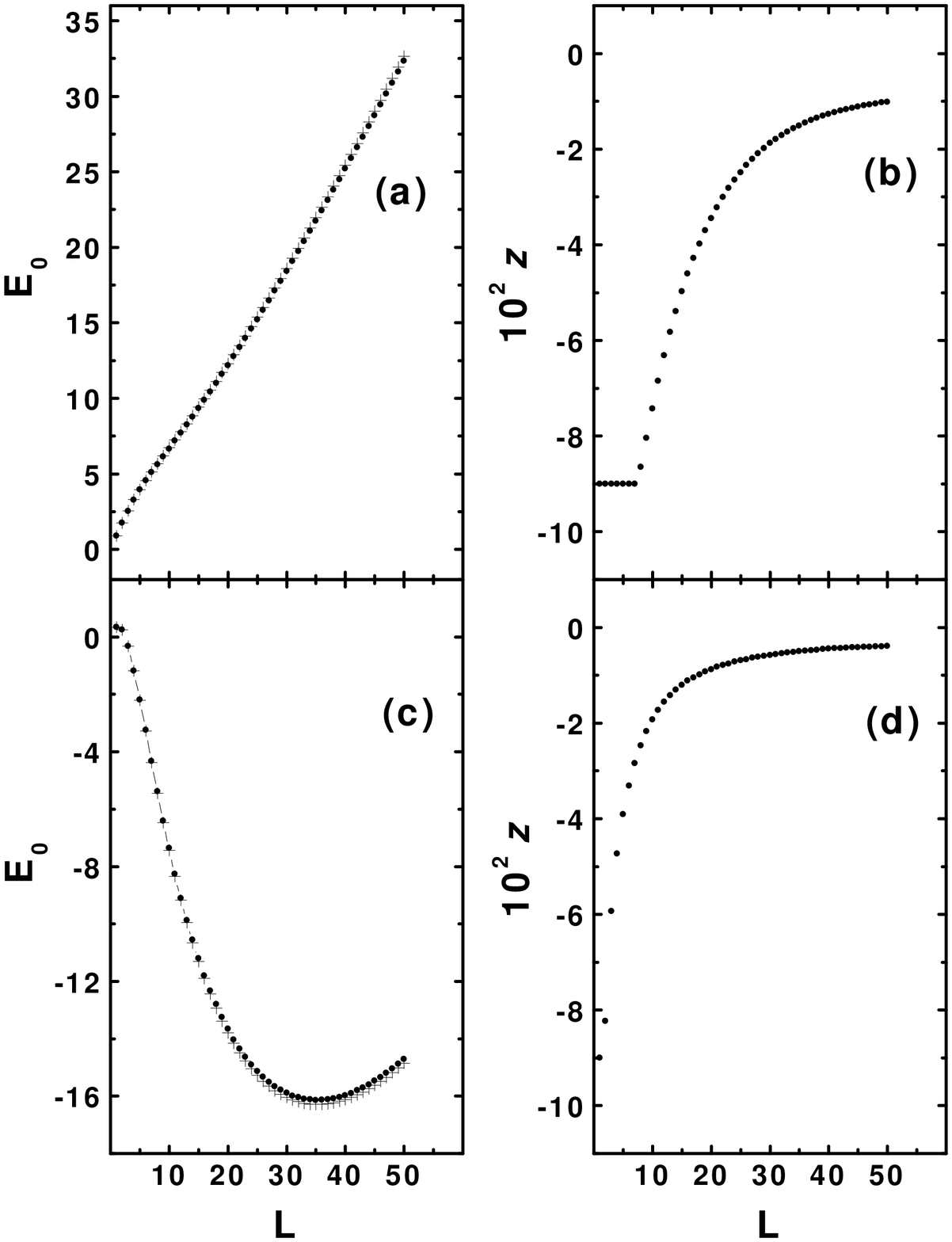}
\end{center}
\caption{Ground-state energies and  $z=\ln (q)$ for the LE model
of Eq. (\ref{b1a}), are shown as a function of $L$. Insets (a) and
(b) show results for the case $N=2 \Omega=30$, $\omega_f=1$ MeV,
$\omega_b=1$ MeV, and $x=G \frac{4 \Omega} { \sqrt{\omega_f
\omega_b}}=0.5$, while insets (c) and (d) correspond to $x=1.5$.
The exact ground state energies corresponding to the LE model of
Eq. (\ref{b1a}) are denoted by crosses while the one corresponding
to its associated effective $su_q(2)$ Hamiltonian of Eq.
(\ref{ca}) are denoted by circles. } \label{fig:fig5}
\end{figure}

\begin{figure}
\begin{center}
\includegraphics[width=10cm]{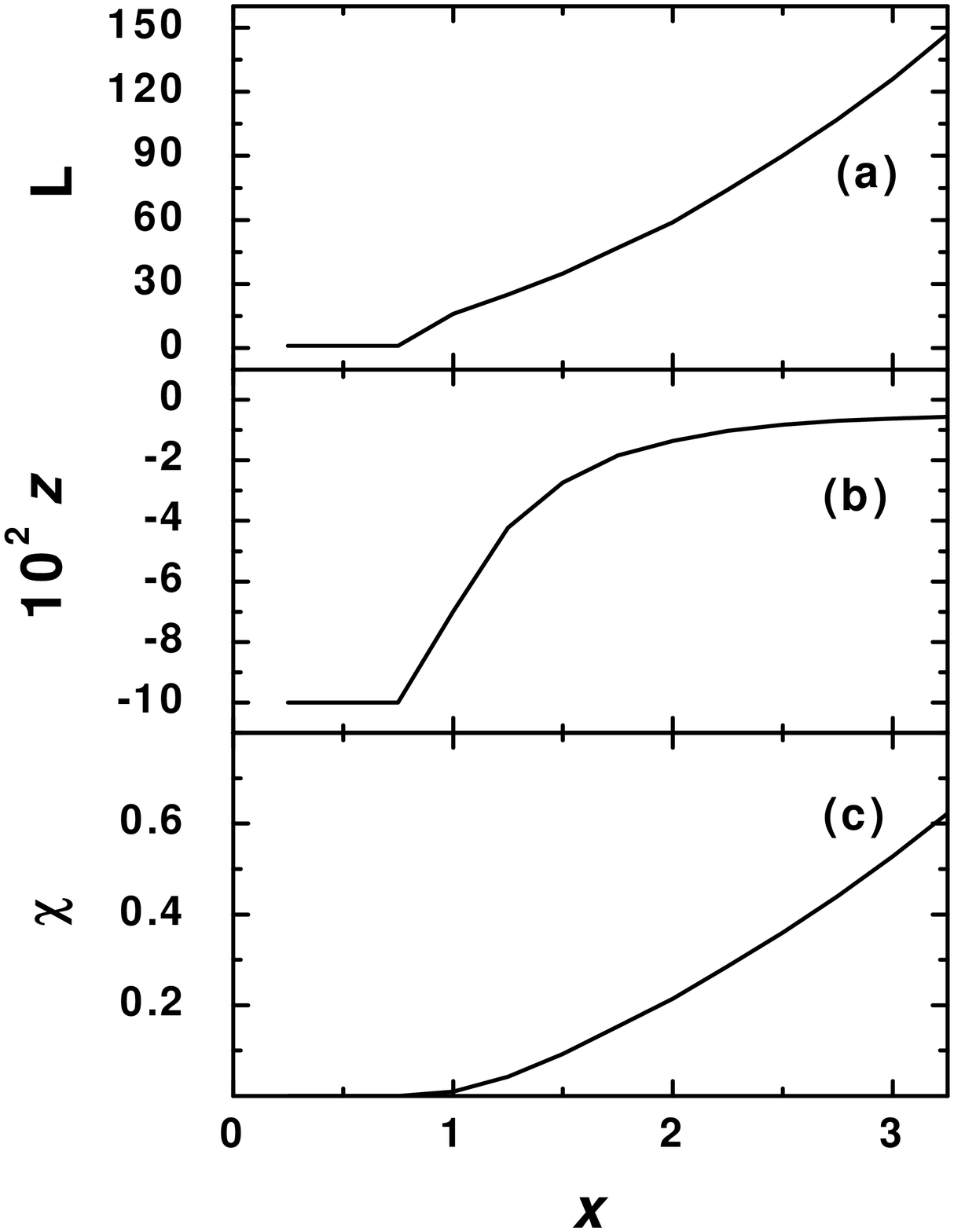}
\end{center}
\caption{Idem as Figure 2, for the LE model of
Eq.(\ref{b1a}), and for the $su_q(2)$ Hamiltonian of Eq.
(\ref{ca}). The values of the parameters $N$, $\omega_f$ and
$\omega_b$ are the same as those given in the captions to Figure
5. } \label{fig:fig6}
\end{figure}

\begin{figure}
\begin{center}
\includegraphics[width=10cm]{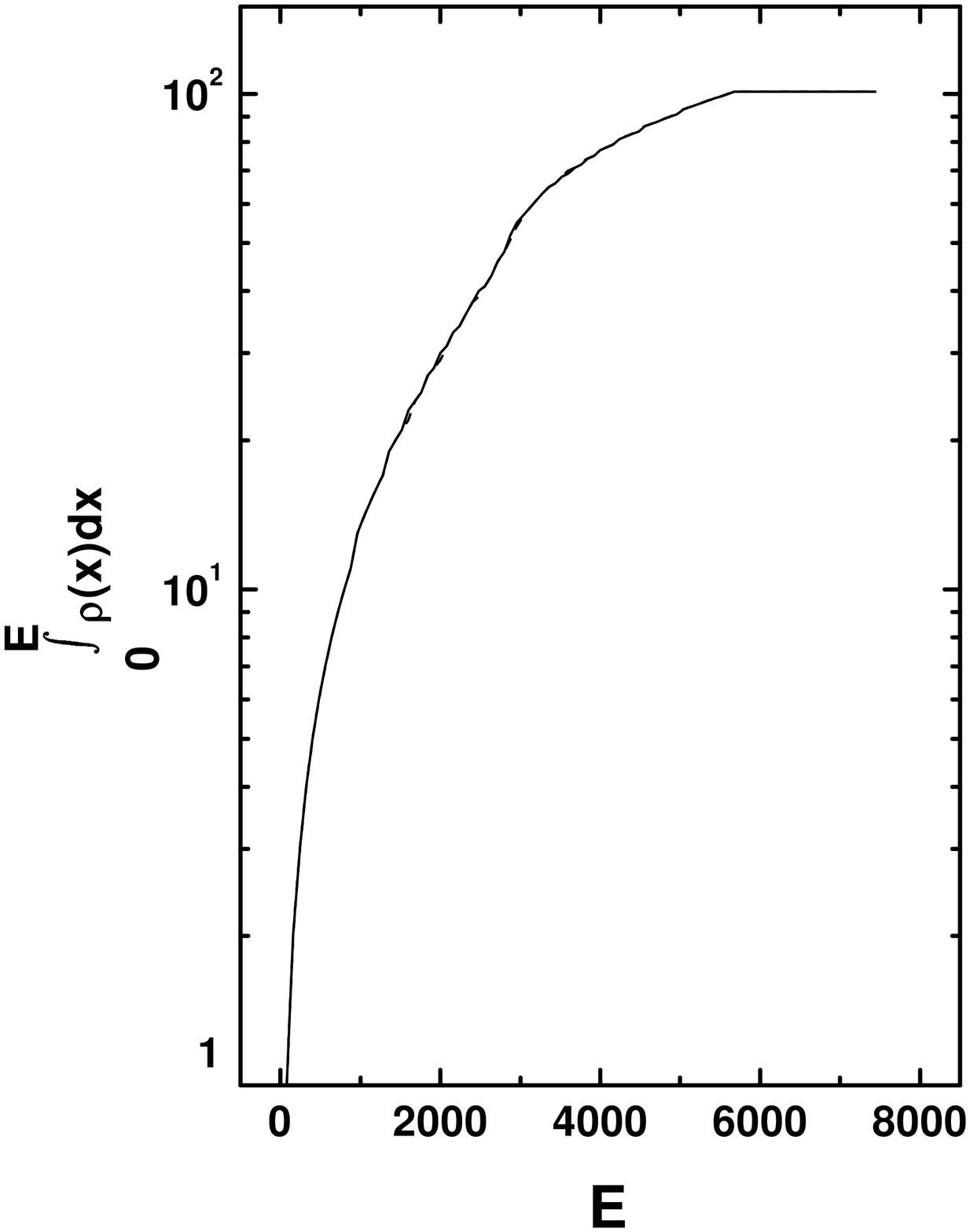}
\end{center}
\caption{ Integrated density of states, for the LE model of Eq.
(\ref{b1a})(solid line) and for the effective $su_q(2)$
hamiltonian of Eq.~(\ref{ca})(dashed line). The values of $N$,
$\omega_f$ and $\omega_b$ are the same as those given in the
captions to Figure 4. The results corresponding to the hamiltonian
of Eq.~(\ref{b1a}) have been obtained with $\frac{L}{2\Omega}=7$.
For the effective $su_q(2)$ hamiltonian of Eq.~(\ref{ca}) the
value $z=- 0.0009$. Both curves coincide within the resolution of
the figure. } \label{fig:fig7}
\end{figure}

\begin{figure}
\begin{center}
\includegraphics[width=10cm]{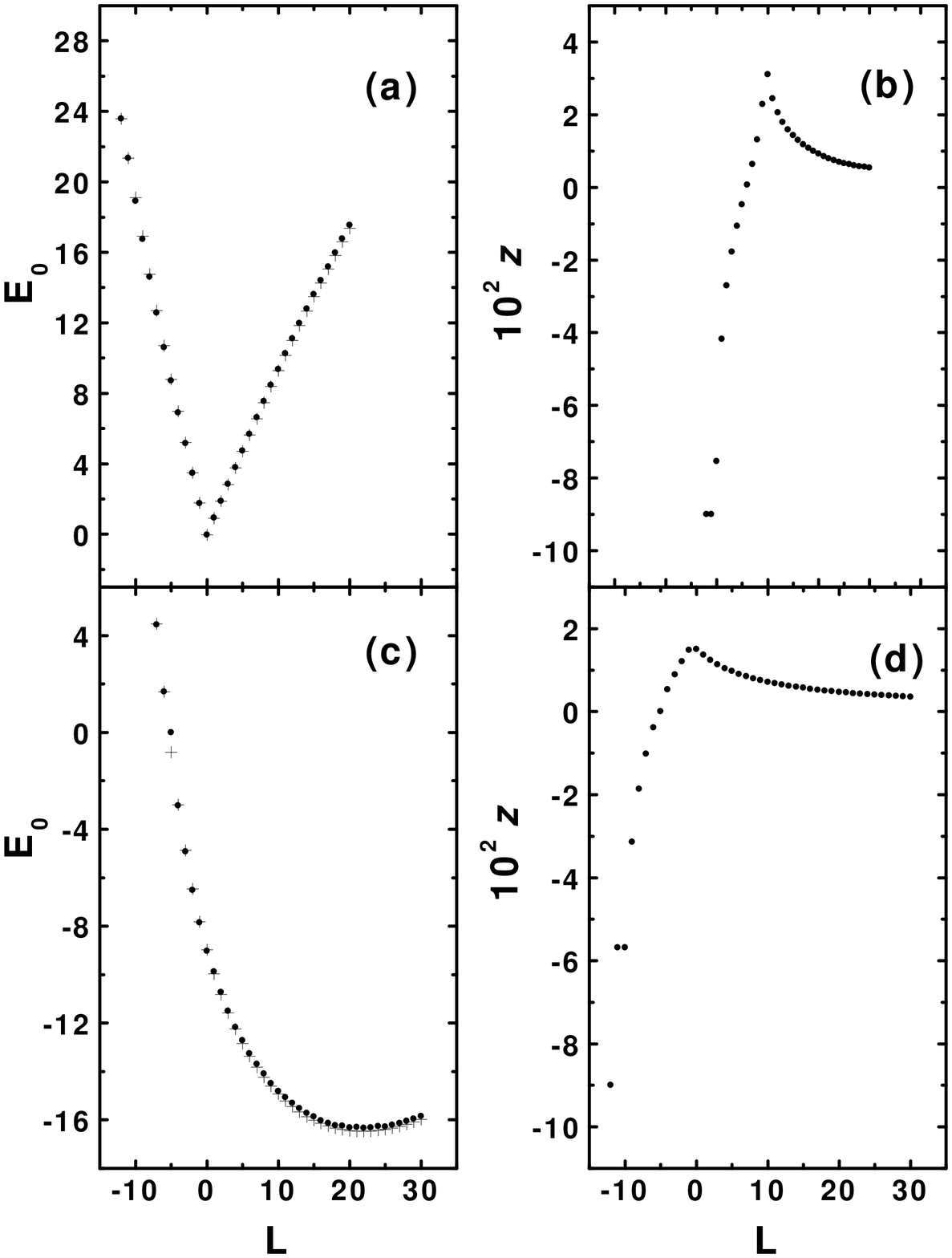}
\end{center}
\caption{Ground-state energies and  $z=\ln (q)$ for the LE model
of Eq. (\ref{b2a}), are shown as a function of $L$. Insets (a) and
(b) show results for the case $N=2 \Omega=30$, $\omega_f=1$ MeV,
$\omega_b=1$ MeV, and $x=G \frac{4 \Omega} { \sqrt{\omega_f
\omega_b}}=0.5$, while insets (c) and (d) correspond to $x=1.5$.
The exact ground state energies corresponding to the LE model of
Eq. (\ref{b2a}) are denoted by crosses while the one corresponding
to the associated effective $su_q(2)$ hamiltonian of Eq.
(\ref{cf}) are denoted by circles. } \label{fig:fig8}
\end{figure}

\begin{figure}
\begin{center}
\includegraphics[width=10cm]{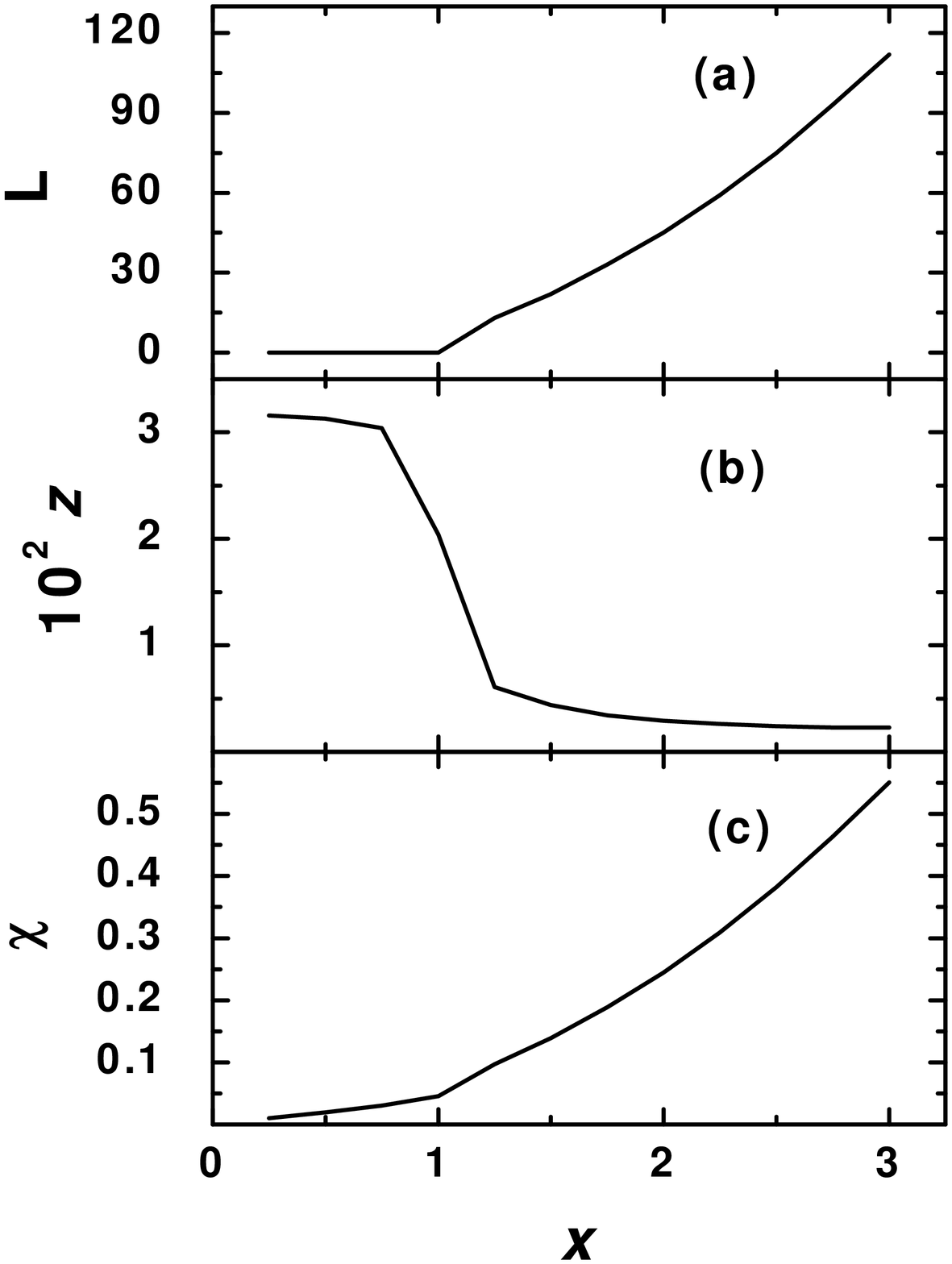}
\end{center}
\caption{Idem as Figure 2, for the  LE model of Eq. (\ref{b2a})
and its correspondent q-analogue of Eq. (\ref{cf}). The values of
$N$, $\omega_f$ and $\omega_b$ are the ones given in the captions
to Figure 8. } \label{fig:fig9}
\end{figure}

\begin{figure}
\begin{center}
\includegraphics[width=10cm]{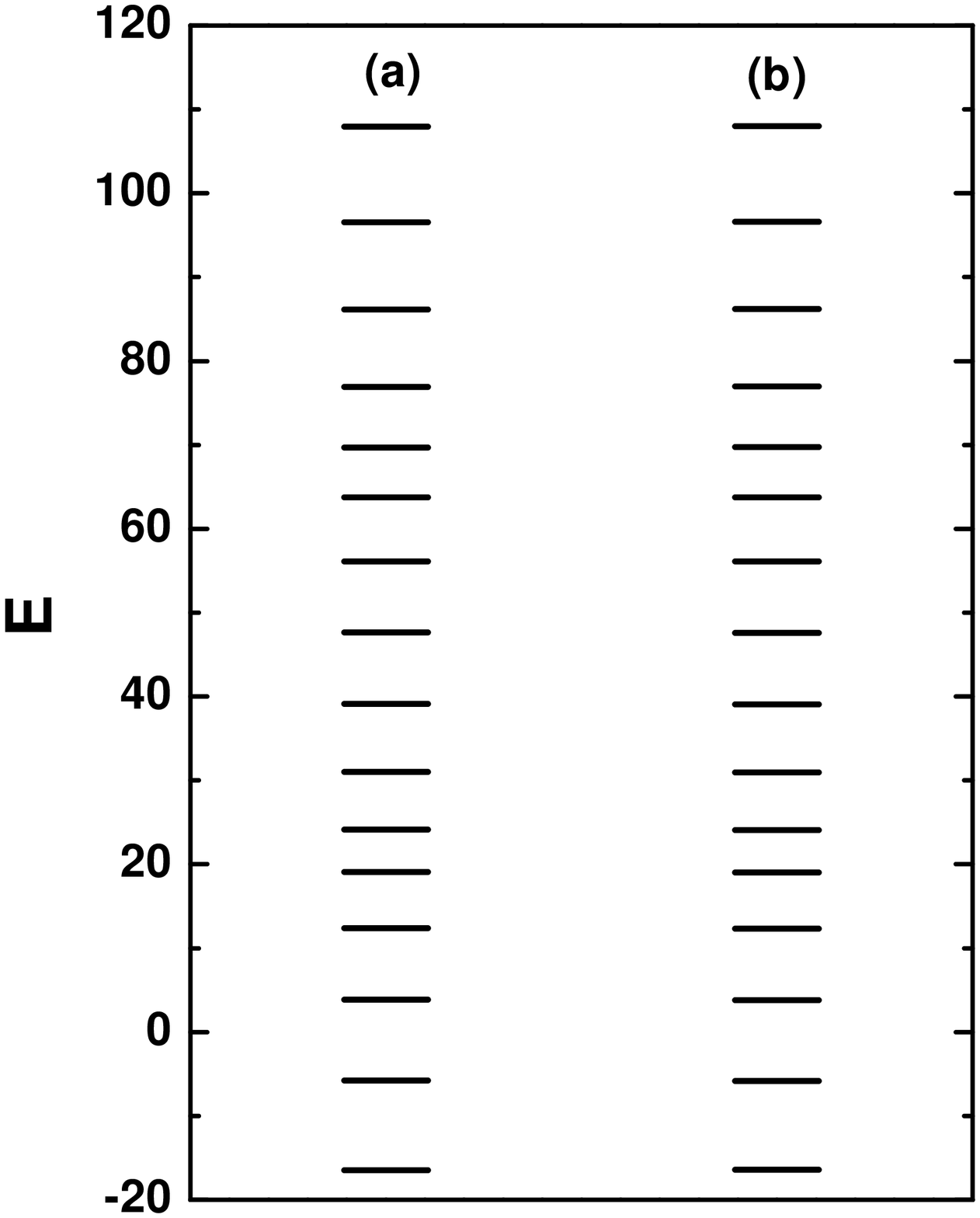}
\end{center}
\caption{The spectrum for the LE model of Eq. (\ref{b2a}) and for
the $su_q(2)$ hamiltonian of Eq. (\ref{cf}). The values of $N$,
$\omega_f$ and $\omega_b$ are the same as those of Figure 8. The
spectrum denoted by (a) corresponds to the one obtained from the
hamiltonian of Eq.~(\ref{b2a}), for $L=22$. The spectrum denoted
by (b) is obtained from the effective $su_q(2)$ hamiltonian of
Eq.~(\ref{cf}), with $z=0.0044$.   } \label{fig:fig10}
\end{figure}

\end{document}